# Broad Angle Resolver for THz Band

Yasith Amarasinghe[1], Yaseman Shiri[2], Hichem Guerboukha[3], Rabi Shrestha[2]
Pernille Klarskov[1] and Daniel M. Mittleman[2]

[1]Department of Electrical and Computer Engineering, Aarhus University, Finlandsgade 22, DK-8200, Aarhus N, Denmark
[2]School of Engineering, Brown University, 184 Hope St., Providence RI 02912 USA
[3]School of Science and Engineering, University of Missouri-Kansas City, 801 E 51 St., Kansas-City MO 64110 USA
*Corresponding author: yasith@ece.au.dk

**Abstract:** Terahertz (THz) communication systems hold immense potential for high-speed data transfer across various domains yet face challenges due to directionality constraints because of free space path loss. To address this, directional beams are commonly employed in THz technology. With the usage directional beams, it is important to track the transmitting device to keep the link connectivity. Leaky Parallel Plate Waveguides (LPPWs) were introduced to tackle this challenge. But traditional LPPW uses unique angle¨-frequency relationship which requires broad range of frequencies. This study proposes a novel approach to mitigate these constraints, balancing directionality with reduced directional gain to enhance LPPW adaptability. The proposed device can accurately determine the receiving angle of a beam by analyzing unique features extracted from the dual peak outputs. Experimentation and simulations reveal that the device allows for a broader angle of acceptance and calculation of the received angle.

## Introduction

Terahertz (THz) systems stand at the forefront of current research and development efforts, showcasing their immense potential for high-speed data transfer across various domains, including wireless communications [1-3] and applications in sensing [4,5] and imaging [6,7]. Despite the promising attributes of THz communications, a challenge arises from the substantial free-space path loss, imposing severe limitations on communication range. To address this challenge there were many solutions introduced to the THz including phased array antennas [8,9], metasurfaces [10-12], graters [13] and Leaky Parallel Plate Waveguides (LPPW) [14-17]. Compared to all the solutions LPPW has the advantage of simple design, lower cost and ability handle wide range of frequencies. LPW has been used in many areas including sensing [16], imaging [18] and multiplexing[14,15]

The conventional design of LPPWs, with a fixed plate separation, introduces a unique characteristic – a frequency-dependent angular peak. While this feature proves advantageous in specific applications, such as frequency division multiplexing [14,15] and sensing[16] it poses a notable challenge when LPPWs are employed as receivers. Since each frequency can only couple into the waveguide from one angle, this will require precise angular alignment between transmitting and receiving antennas. The plate separation also has to be precise as few micrometer difference in them will result in few degrees of change at the coupling angle.

This paper aims to improve this terahertz communication systems by addressing this key challenge: stringent directionality constraints in LPPWs. We propose a novel approach that remove this precise angular alignment constraint with a reduction in directional gain.. In traditional LPPW,

the THz beam couples into and out of a rectangular slot following the frequency-angle relationship [14]

$$f(\theta) = \frac{mc_0}{2b\sin(\theta)} \quad (1)$$

where $c_o$ is the speed of light, m is the TE mode order and b is the plate separation of the LPPW. From this equation, we can see that the coupling angle ($\theta$) can be manipulated by changing the plate separation. If we keep the operating frequency as a constant we can manipulate the coupling angle as a variable to the plate separation (fig. 1(a)). For our proof-of-concept experiment, we choose an operating frequency of 200 GHz. The coupling angle ($\theta$) and plate separation relation for $TE_1$ and $TE_2$ modes at 200 GHz are shown in Fig. 1(a).

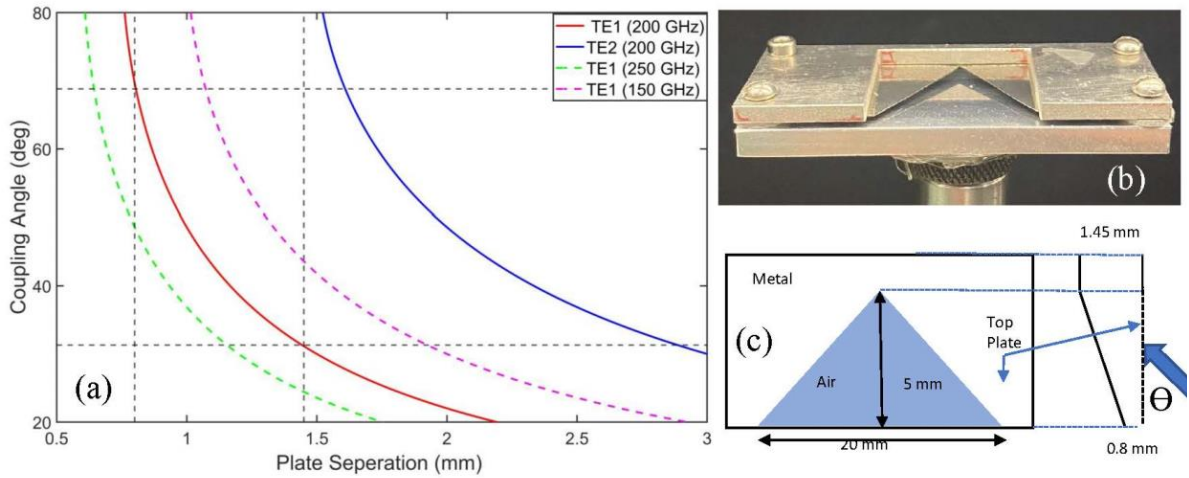

**Fig**. 1. (a) Plate separation vs coupling angle ($\theta$). (b) Actual device. (c) Top view of the device and corresponding side profile of the device.

Another disadvantage of the LPPW is the cut off frequency which will restrict the usable frequency range. The cut off frequency of the LPWG is given by $f_c = mc_0/2b$. But in this experiment, we use cutoff frequency to restrict higher order modes(i.e., exciting the $TE_2$ mode) from inducing inside the LPPW. We choose the plate separation to change from 0.8 mm to 1.45 mm (Fig. 1(c)). The $2^{nd}$ order mode for 200 GHz will start inducing inside the LPPW after 1.5 mm plate separation. Theoretically, this allows the receiver to work as a broad angle receiver of angles from 31° to 63° at 200 GHz. We also observe that this device satisfy some coupling angle for lower and higher frequencies. This opens the possibility of this device not only working as a single frequency device, but a broadband device as well, so long as it satisfies the coupling angle to plate separation equation.

## Design and Fabrication

is device created using two metallic plates put together with the one of the plates slightly tilted to produce a linear change of the plate separation. The top plate has triangular opening which work as the slot for the LPPW as shown in the actual device in Fig. 1 (b).. The top plate view and corresponding side plate profile are shown in Fig. 1 (c). The receiving beam is coupled into the device at an angle Θ as shown in the same figure. The plate separation varies from 0.8 mm in the higher triangular opening to 1.45 mm at the end of the triangular opening where the receiver antenna is held. This triangular slot geometry with a width that linearly increases along its length Fig. 1(c).In a traditional LPPW, this triangle slot can preserve the frequency-angle relation while enhancing output coupling efficiencies and narrowing the radiated beam's angular directions, hence increasing the beams directionality [17]. This triangular slot also utilizes the whole beam coupled into the LPPW, which increases the efficiency of the beam compared to rectangular slot. Although we are using a triangular slit and utilizing the whole beam, it will not increase the directionality as the plate separation changes inside the LPPW. Instead it will use slice of the beam which will match the plate separation for the particular incoming angle and coupled in from both sides of the device. We performed finite element method simulations (Fig. 2 (a)) on the above-mentioned device as a receiver and observed that the intended different plate separations can couple different angles with the operation frequency of 200 GHz. When the coupling angle is smaller (around 30º), the device can couple the beam at the smaller triangular opening where the plate separation is larger. When the coupling angle is larger (greater than 50º), the device can couple the beam at the two larger openings of the device where the plate separation is closer to 0.8 mm. This is clearly shown in the numerical simulation results (fig. 2 (b)).

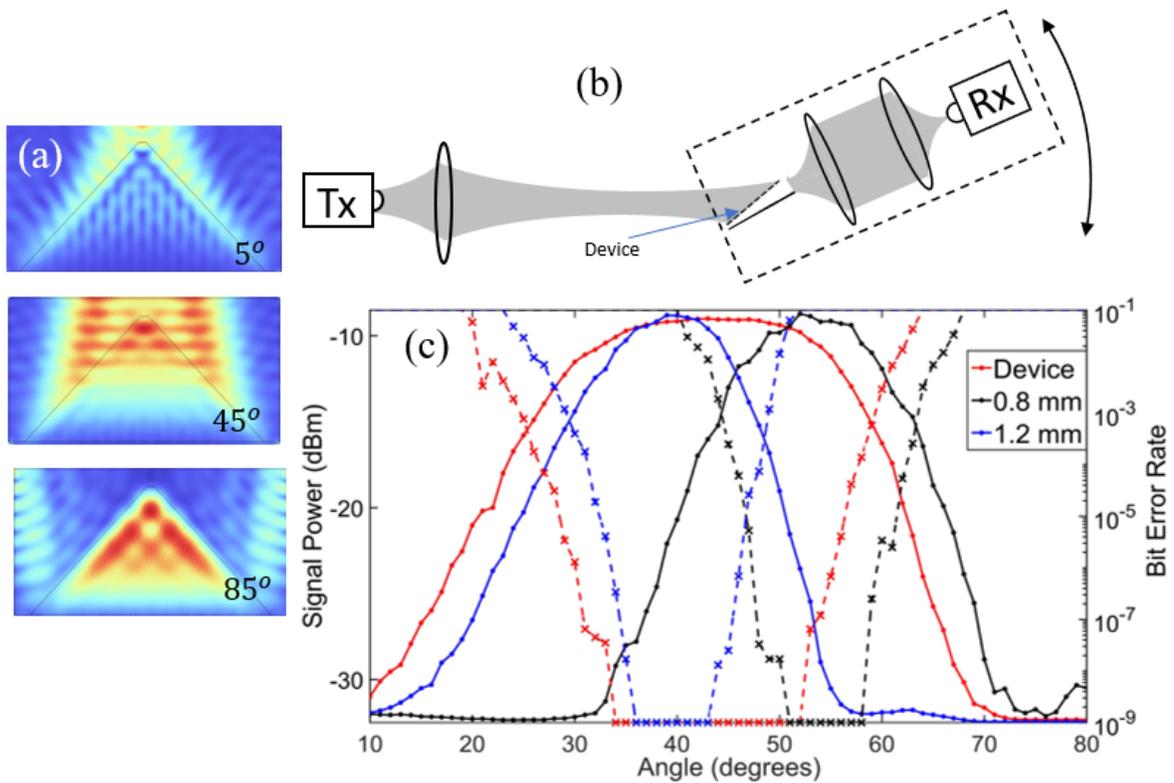

**Fig. 2** (a) Simulation of beam path inside the device for coupling angles of 5º,45º and 85º. (b) Experimental setup for wide angle measurements. (c) Received Signal power vs coupling angle (solid line) and corresponding BER vs coupling angle (dashed line).

## Experiment

Next, we experimentally characterize the device using an on-off keying (OOK) modulated communications system. The experimental setup (Fig. 2 (b)) uses a 1 Gbit per second data rate system to send data through the THz beam. We measured the efficiency of our device against normal LPPW [14] using both received power and bit error rate (BER). The receiver part of the setup (in the dotted rectangle (Fig. 2 (a)) rotates with the device as the center, with the received power and BER measured for each 1° rotation. The results are shown in fig. 2 (c). As can be seen by the red curve, this device possesses a large receiving angular aperture (with a half-power beamwidth of 27°). We repeat the experiment with two reference devices that have constant plate separations of 0.8 mm and 1.2 mm (shown as black and blue curve respectively). As expected from Eq. 1, the antenna with 0.8 mm plate separation peaks at 52° degrees, while the antenna with 1.2 mm plate separation peaks at 39° degrees. In these cases, we obtain half-power beamwidths of 12° and 14°, which indicate that our proposed device has a broader angular acceptance than either of the two reference waveguides. But the device acceptance angle was smaller than the theoretical angle mentioned in the introduction. This is because at higher angles the beam couples into the device the edge of the triangle slot and slips from the horn antenna which is situated at the center. By further analyzing the output profile of the simulation results (Fig. 3 (a)), we confirm that the output exhibited two distinct peaks (fig. 2 (a)) for higher input angles, instead of a single peak. This happens because both sides of the triangle slot (fig. 2(b)) meet the criteria for plate separation-coupling angle relation mentioned in Eq. (1). Therefore, the beam will couple into the device from both sides at same time, which will translate to two distinct peaks.

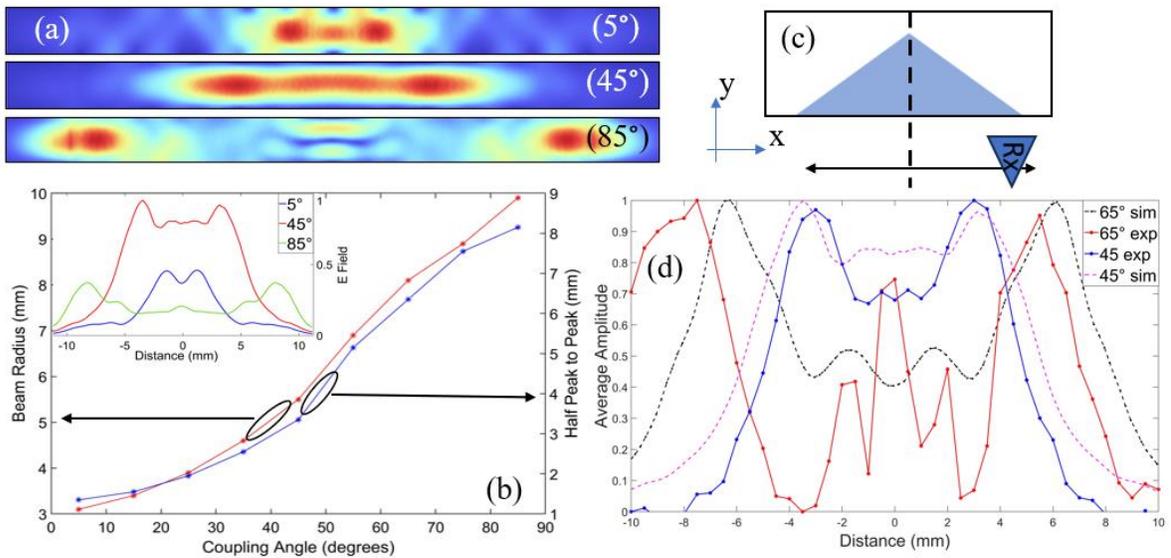

**Fig. 3** (a) Simulation results for E field norm of coupling angles 5°, 45° and 85°. (b) Simulation results of beam radius of the output and half peak-to-peak distance of the output vs coupling angle (b-inset) Simulation results for E Field intensity along the output of the device. (c) Experimental setup for linear scan at the receiver's end. (d) Comparison of simulation and experiment results at the output for average amplitude vs. distance from the axis.

One of the disadvantage of this device compared to the traditional LPPW device was we couldn't estimate the arriving angle because we are accepting the same frequency with a broader angle. But, if our device can produce unique signatures for different angles we could match these unique features of the receiving beam to identify the receiving angle. Since the triangular slot width is increasing linearly, we theorize that the distance between the two distinct peaks produced by the output of the device also increases linearly. To validate these findings, we conducted simulations of the device's output profiles across various coupling angles. In these simulations, a planar wave was coupled into the triangular opening, and the resulting output electric field was observed at the device output. Fig. 3(a) displays the electric field norms for three different outputs. At higher angles, we observed the emergence of higher-order modes induced in the center, alongside the presence of two main lobes. To quantify our observations, we integrate the electric field across the $z$ direction and plot the integrated electric field versus distance from the center in $x$ direction. Two beam profiles according to this method are shown in Fig 3 (d) for two different input angles. We can observe that at lower angles the two peaks are very close to each other, and the power of the beam is concentrated at the center. However, a larger input angle results in the two peaks becoming separated. At much higher angles, they become two completely separate beams. We can measure the beam radius of the entire beam as a single entity, as well as the distance between the two peaks across different angles. We established a clear relationship between the incident wave's angle and the beam radius, as depicted in Fig. 3(b). Both the beam radius and the distance between the two peaks can be used to identify the coupling angle into the device.

To further verify our findings, we conducted an experiment using the same setup as fig. 2 (b). But the output is scanned linearly as shown in fig. 3 (c). It is important to note that at higher angles (65⁰) we could clearly see the two peaks as they are well separated. But when we try to conduct the same experiment with a lower angle (45⁰) with the same horn antenna we couldn't identify two distinct beams as the two peaks are close to each other and horn antenna captures both of the peaks at the same time. In order to increase the resolution of our linear scan we put two aluminum plates across the horn antenna surface to reduce the opening (fig. 2 (b-inset)). The experimental data closely match with the simulation data as shown in Fig. 3(c). These experimental data confirm our ability to accurately determine the angle of arrival using our broad-angle receiver. Given that the main beam splits into two beams at higher angles, optimal reception may be achieved with a larger antenna at the receiver at the cost of calculation of the angle of arrival or an array of smaller antennas to capture both lobes effectively.

## Conclusion

Our study proposes a novel approach to address the directionality constraints of Leaky Parallel Plate Waveguides in terahertz communication systems. By balancing directionality with a reduction in directional gain, we enhance the adaptability of LPPWs, particularly in scenarios where achieving perfect angular alignment is challenging. Through experimentation and simulation, we demonstrate that maintaining the receiver antenna at the center allows for a broader angle of acceptance, although lower than initially anticipated. Our findings reveal the emergence of two distinct peaks in the output profile at higher angles attributed to both sides of the device satisfying the coupling angle to plate separation criteria. Moreover, we establish clear markers for identifying the coupling angle into the device, facilitating accurate determination of the angle of arrival. This research significantly advances the field of THz communications, making leaky-wave devices more versatile and user-friendly, with implications extending beyond wireless communication to sensing and imaging applications.